\documentclass[12pt]{iopart}

\usepackage{iopams}  
\usepackage{graphicx}

\begin{document}

\title[Collisionless discontinuity]{Three-dimensional structure and stability of discontinuities between unmagnetized pair plasma and magnetized electron-proton plasma}

\author{M. E. Dieckmann}
\address{Dept. of Science and Technology (ITN), Link\"oping University, Campus Norrk\"oping, SE-60174 Norrk\"oping, Sweden}
\ead{mark.e.dieckmann@liu.se}
\author{D.~Folini}
\address{Univ Lyon, ENS de Lyon, Univ Lyon 1, CNRS, Centre de Recherche Astrophysique de Lyon UMR5574
F-69230, Saint-Genis-Laval, France}
\author{M.~Falk}
\address{Dept. of Science and Technology (ITN), Link\"oping University, Campus Norrk\"oping, SE-60174 Norrk\"oping, Sweden}
\author{A.~Bock}
\address{Dept. of Science and Technology (ITN), Link\"oping University, Campus Norrk\"oping, SE-60174 Norrk\"oping, Sweden}
\author{P.~Steneteg}
\address{Dept. of Science and Technology (ITN), Link\"oping University, Campus Norrk\"oping, SE-60174 Norrk\"oping, Sweden}
\author{R.~Walder}
\address{Univ Lyon, ENS de Lyon, Univ Lyon 1, CNRS, Centre de Recherche Astrophysique de Lyon UMR5574
F-69230, Saint-Genis-Laval, France}

\vspace{10pt}
\begin{indented}
\item[]August 2022
\end{indented}

\begin{abstract}
We study with a 3D PIC simulation discontinuities between an electron-positron pair plasma and magnetized electrons and protons. A pair plasma is injected at one simulation boundary with a speed 0.6$c$ along its normal. It expands into an electron-proton plasma and a magnetic field that points orthogonally to the injection direction. Diamagnetic currents expel the magnetic field from within the pair plasma and pile it up in front of it. It pushes electrons, which induces an electric field pulse ahead of the magnetic one. This initial electromagnetic pulse (EMP) confines the pair plasma magnetically and accelerates protons electrically. The fast flow of the injected pair plasma across the protons behind the initial EMP triggers the filamentation instability. Some electrons and positrons cross the injection boundary and build up a second EMP. Electron-cyclotron drift instabilities perturb the plasma ahead of both EMPs seeding a Rayleigh-Taylor-type instability. Despite equally strong perturbations ahead of both EMPs, the second EMP is much more stable than the initial one. We attribute the rapid collapse of the initial EMP to the filamentation instability, which perturbed the plasma behind it. The Rayleigh-Taylor-type instability transforms the planar EMPs into transition layers, in which magnetic flux ropes and electrostatic forces due to uneven numbers of electrons and positrons slow down and compress the pair plasma and accelerate protons. In our simulation, the expansion speed of the pair cloud decreased by about an order of magnitude and its density increased by the same factor. Its small thickness implies that it is capable of separating a relativistic pair outflow from an electron-proton plasma, which is essential for collimating relativistic jets of pair plasma in collisionless astrophysical plasma.
\end{abstract}

%
\vspace{2pc}
\noindent{\it Keywords}: Plasma discontinuity, Rayleigh-Taylor instability, Magnetic flux ropes, Pair plasma, PIC simulation, Astrophysical Jet

\submitto{\NJP}
%
\maketitle
%
%

\section{Introduction}

Microquasars, which are binary systems formed by a regular star and a compact object that can be a neutron star or a black hole, can give rise to relativistically fast jets~\cite{Mirabel1994,Carotenuto2021}. Microquasars are powered by magnetic extraction of angular momentum from the rotating compact object~\cite{BlandfordZnajek1977} or from the accretion disk~\cite{BlandfordPayne1982}, which surrounds the compact object and is filled with material captured from the regular star. The conditions in the region between the compact object and the inner accretion disk are extreme (See Ref.~\cite{Yuan2014} for a recent review). Intense electromagnetic radiation reaching well into the hard X-ray band and the swirling of ultrastrong electromagnetic fields ionize and heat the material of the accretion disk and its surrounding corona. Hot ionized gas is known as plasma. It reaches temperatures that let friction caused by binary collisions between particles become negligible compared to the electromagnetic forces induced by the electric current due to the collective motion of charged particles. 

The coronal plasma, which surrounds the inner part of the accretion disk, can get so hot that X-rays and plasma particles reach energies above the rest energy of an electron-positron pair. Such conditions trigger pair production cascades, which let macroscopic clouds of pair plasma form. Observations of pair annihilation lines during X-ray flares of microquasars are evidence of their presence~\cite{Siegert2016}. Microquasar jets have also been named as a possible source of galactic positrons~\cite{Prantzos2011}.

Gradients of the thermal and magnetic pressure, radiation, and moving magnetic fields accelerate the plasma particles. Some particles escape from the vicinity of the accreting compact object. The density of this outflow and that of ambient material, into which it expands, are so low that the plasma particles can travel enormous distances without colliding with other particles; yet there must be a mechanism that channels some of this outflow into a jet. Collimation requires a discontinuity~\cite{Roth1996,Myong1997} that separates the outflow from the surrounding ambient material. If it exists, then the outflow acts as an expanding bubble that expels the ambient material. The bubble expands faster along the outflow's mean flow direction than along the other directions, where the expansion is driven by thermal pressure, giving the jet its pencil-shaped form. Despite their importance for astrophysical jets, plasma discontinuities between pair plasma and ambient electron-ion plasma have not received much attention in the past.

We examine with the EPOCH particle-in-cell (PIC) code~\cite{Arber2015} the evolution of two plasma discontinuities in physically realistic three-dimensional space. Both plasma discontinuities grow self-consistently in the layer between an unmagnetized pair plasma, which represents the outflow and is injected at a boundary, and ambient magnetized electron-proton plasma. They are pushed into the ambient plasma by pair plasma, which has a low mean speed and is close to a thermal equilibrium behind one discontinuity and is fast-flowing behind the other. The expanding pair plasma expels the magnetic field of the ambient plasma and piles it up at its front. The magnetic pulse traps ambient electrons and pushes them across the protons. Their current induces an electric field, which changes the initial magnetic pulse into an electromagnetic pulse (EMP). Both EMPs confine the pair plasma magnetically and accelerate the protons electrically, which reduces the proton density in the volume occupied by the pair plasma and lets the EMPs act as discontinuities. The discontinuity that confines the quasi-thermal pair plasma remains compact during the simulation time. The one in front of the fast-moving pair plasma is a broad transition layer, which is nevertheless capable of slowing down the expansion of the pair plasma and accelerating the protons on spatial scales that are small compared to the spatial scales of astrophysical jets.

Section~2 in our paper presents relevant hydrodynamic structures and their counterparts in collisionless plasma, the PIC method, and the simulation setup. Section~3 presents our results. Their significance is discussed in Section~4.

\section{Related work}

\subsection{Hydrodynamic jet model}
\begin{figure}[ht]
\includegraphics[width=\columnwidth]{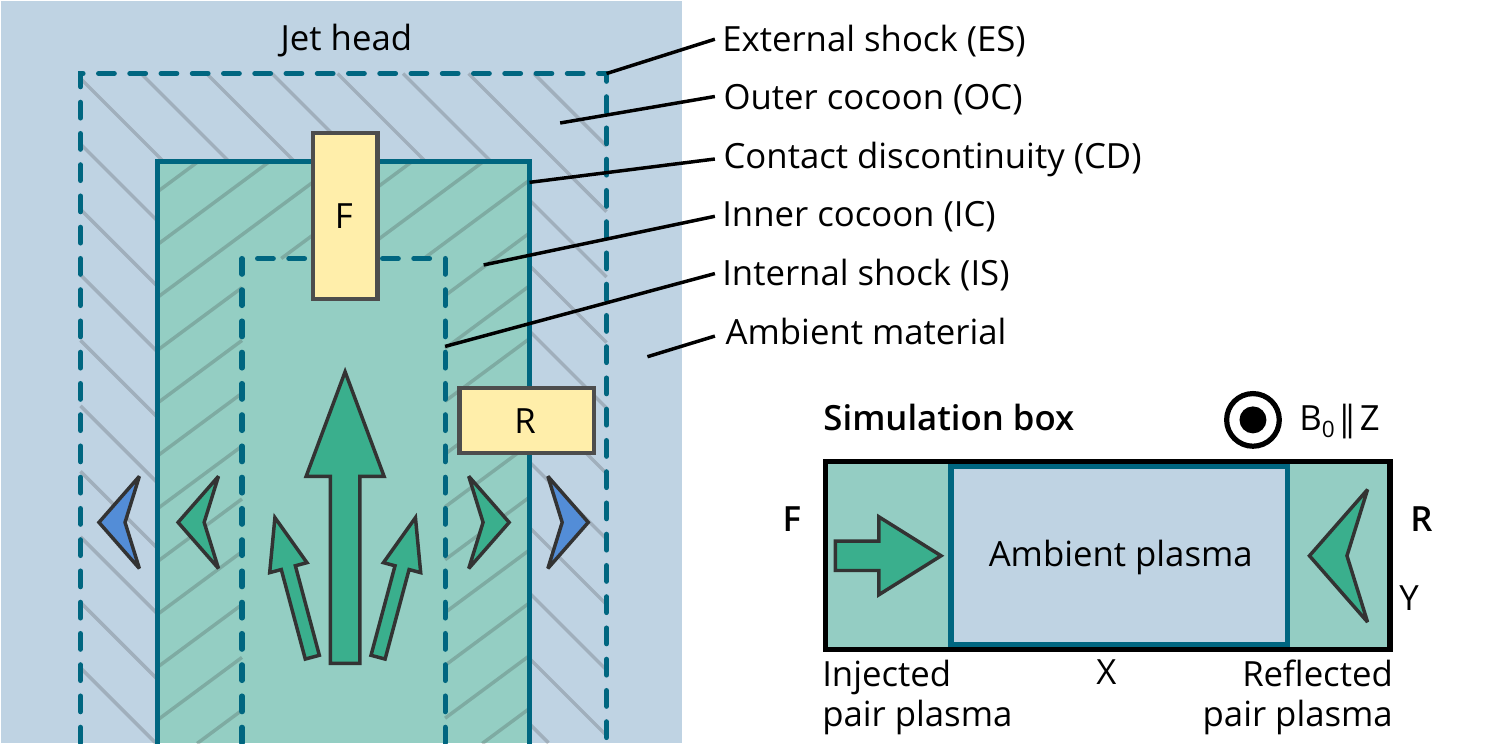}
\caption{The cross-section of a hydrodynamic relativistic jet and our simulation setup are sketched to the left and right. The pair outflow and the structures in it are shown in green, while the ambient electron-proton plasma and the structures immersed in it are colored blue. The mean velocity vector of the outflow points upwards (green arrows). Arrowheads indicate lateral thermal pressure-driven plasma motion. The CD separates the shocked ambient material of the OC from the shocked outflow material of the IC. In our simulation, we use periodic boundaries and inject a pair plasma at the left boundary that expands across a perpendicular background magnetic field. We give the pair plasma a mildly relativistic speed to the right (green arrow), which gives conditions similar to those in box F (forward) in the jet model. Its interaction with the ambient plasma scatters and heats the pair plasma. Some of it returns to the injection boundary, crosses it, and expands from the right boundary to the left. This expansion is driven by the thermal pressure of the scattered pair plasma (green arrowhead), which is realistic for the plasma in box R (reflected) in the jet model.}
\label{figure1}
\end{figure}

In a collisional hydrodynamic jet model~\cite{Aloy1999,Bromberg2011,Charlet2022}, a contact discontinuity maintains the separation of the outflow from the surrounding gas giving rise to the structure sketched in Fig.~\ref{figure1}.
The relativistically fast outflow is stopped by the discontinuity, which compresses and heats it. The region filled with slowly moving, hot, and dense outflow material is called the inner cocoon. If the mean speed of the freely moving outflow material relative to that of the inner cocoon exceeds the sound speed, the boundary of the inner cocoon changes into an internal shock. The thermal pressure of the heated outflow material pushes the contact discontinuity outwards into the surrounding ambient material, which is the stellar wind of the companion star or the interstellar medium. The ambient material is set in motion and heated by the expanding contact discontinuity and an outer cocoon forms. It is bounded by an external shock if the expansion speed of the outer cocoon exceeds the speed of sound in the ambient material. 

\subsection{Collisionless plasma and its numerical approximation}

The equations, which are solved by PIC simulation codes, can resolve all waves and structures found in collisionless plasma and can be normalized by selecting characteristic scales for space and time. The proton plasma frequency $\omega_{pi}={(e^2n_0/\epsilon_0 m_p)}^{1/2}$ normalizes time, where $e$, $m_p$, $\epsilon_0$, and $n_0$ are the elementary charge, the proton mass, the vacuum permittivity, and the proton number density. We normalize space with the proton skin depth $\lambda_{pi} = c/\omega_{pi}$ ($c$: speed of light in vacuum). The amplitudes of the electric field $\mathbf{E}(\mathbf{x},t)$, of the magnetic field $\mathbf{B}(\mathbf{x},t)$, and the macroscopic current density $\mathbf{J}(\mathbf{x},t)$ are defined on a spatial grid and normalized to $\omega_{pi}m_pc/e$ and $\omega_{pi}m_p/e$, and $c e n_0$. In this normalization, the \textit{EPOCH} code evolves the electromagnetic fields in time with discretized forms of  Amp\`ere's law and Faraday's law
\begin{equation}
\nabla \times \mathbf{B} = \frac{\partial \mathbf{E}}{\partial t} + \mathbf{J}, \hskip 2cm \nabla \times \mathbf{E} = -\frac{\partial \mathbf{B}}{\partial t}. 
\end{equation}
The numerical scheme of the \textit{EPOCH} code is that proposed by Esirkepov~\cite{Esirkepov2001}, which fulfills Gauss' law and $\nabla \cdot \mathbf{B} = 0$ to round-off precision. Each plasma species $i$ is approximated by a set of computational particles (CPs). The charge $q_j$ and mass $m_j$ of the $j^{th}$ CP that represents species $i$ with $q_j/m_j = q_i / m_i$ are normalized to $e$ and $m_p$. Its relativistic momentum $\mathbf{p}_j = m_i \gamma_j \mathbf{v}_j$ ($\gamma_j$: relativistic factor) is normalized to $m_p c$. The macroscopic current density $\mathbf{J}$ is obtained from the sum of the contributions of all CPs. The momentum of each CP is updated with the relativistic Lorentz force equation 
\begin{equation}
\frac{d\mathbf{p}_j}{dt} = q_j \left (  \mathbf{E}(\mathbf{x}_j) + \mathbf{v}_j \times \mathbf{B}(\mathbf{x}_j)  \right ),
\end{equation}
using the electric and magnetic field amplitudes at the CP's position $\mathbf{x}_j$. Since we can normalize the Maxwell-Lorentz set of equations, the results do not depend on the actual value $n_0$ of the proton density. Once we set the value of $n_0$, we can retrieve physical units by multiplying the normalized quantities with the scaling factors, which depend only on $\lambda_{pi}$, $\omega_{pi}^{-1}$ and physical constants. We did not include radiation reaction and pair annihilation processes in our simulation, which have their own characteristic time scale, in order to maintain the scaling of our results with a single frequency $\omega_{pi}^{-1}$ and $c/\omega_{pi}$.

\subsection{Previous work}

Previous related PIC simulations have focused on the structure of mildly and highly relativistic shocks, which bound the inner and outer cocoons near the jet head in Fig.~\ref{figure1}. References~\cite{Nishikawa2005,Chang2008,Dieckmann2018,Dieckmann2020d} present results from simulations of internal shocks in pair plasma, while external shocks in electron-ion plasma are discussed in Refs.~\cite{Frederiksen2004,Spitkovsky2008}. Reference~\cite{Marcowith2016} is a review. Relativistic shocks are mediated by the current filamentation instability (See Ref.~\cite{Bret2010} for a review), which leads to a strongly magnetized transition layer with a width that exceeds $\lambda_{pi}$ in the case of electron-proton shocks and the electron skin depth $\lambda_{pe}=\lambda_{pi}/ \sqrt{1836}$ in the case of pair plasma shocks. 

PIC simulations have also addressed spatially localized pair plasma outflows in ambient electron-ion plasma~\cite{Dieckmann2019} and spatially localized electron-ion outflows in electron-ion plasma ~\cite{Yao2019} (See Ref.~\cite{Nishikawa2019} for a review). Spatially localized means that the diameter of the outflow perpendicularly to its mean velocity vector was less than the simulation box size in this direction, which is a prerequisite for the formation of a jet. 

Several simulation studies of discontinuities between pair plasma and electron-proton plasma exist. They resolved either the pair cloud's expansion direction~(1D) or this direction and one orthogonal to it~(2D), which assumes that the discontinuity is planar and of infinite extent in the unresolved directions. One mechanism that can establish a discontinuity between a pair plasma and an unmagnetized ambient electron-proton plasma is based on the different number densities of positrons and electrons in its transition layer. If the pair cloud pushes the electrons and positrons in this layer, the net current drives the electric field that accelerates protons and positrons. Instabilities between the pair particles and protons can drive ion acoustic shock waves~\cite{Dieckmann2018a,Huang2022}. 

An EMP that separated pair plasma from a magnetized electron-proton plasma grew in a 1D simulation~\cite{Dieckmann2020a}. It was tracked long enough to show that an outer cocoon formed. The EMP is pushed by a pair plasma into an ambient plasma with protons and is thus susceptible to Rayleigh-Taylor (RT) instabilities. Hydrodynamic RT instabilities of discontinuities in astrophysical plasma are discussed in Ref.~\cite{Allen1984}. They can also be found in magneto-hydrodynamic (MHD) fluids. If an initially planar boundary is permeated by a uniform magnetic field, which is aligned with one of its directions, the RT instability involves two separate modes. Undular modes have wavevectors that are parallel to the magnetic field direction while wavevectors of interchange modes are perpendicular to it. Magnetic tension can only limit the growth of the undular mode and the interchange mode is thus more destructive~\cite{Liu2009}. Winske~\cite{Winske1996} derived growth rates of both modes of the RT-type instability of a discontinuity in magnetized collisionless electron-ion plasma. 

PIC simulations found that the EMP is unstable in 2D. When the magnetic field was oriented in the simulation plane and orthogonal to the pair outflow's expansion direction, the EMP was unstable to an RT-type instability~\cite{Dieckmann2020b}. The undular mode deformed the boundary but it could not destroy it. When the background magnetic field pointed out of the simulation plane, the EMP collapsed on a time scale that was short compared to the growth time of the interchange mode~\cite{Dieckmann2022}. The EMP was replaced by a broad transition layer that was nevertheless capable of confining the pair plasma and accelerating protons in the expansion direction of the pair plasma. In both simulations, the transition layer's thickness remained small compared to jet scales. 

A comparison of the 2D simulations with an in-plane and out-of-plane magnetic field evidenced that the true structure of the discontinuity's transition layer can only be resolved by a 3D simulation. According to the 2D simulations, the discontinuity forms during a few $\omega_{pi}^{-1}$ and the instability that changes it involves spatial scales between $\lambda_{pe}$ and $\lambda_{pi}$, which is within reach for a 3D PIC simulation.

\subsection{Initial conditions}

We fill the simulation box, which has periodic boundary conditions in all directions, with a uniform ambient plasma. It consists of electrons with the mass $m_e$ and protons with the mass $m_p =1836m_e$. Both species have the number density $n_0$ and temperature $T_0= 2000$ electronVolts~(eV). For these initial conditions, the normalized thermal speeds of electrons $v_{te}={(k_BT_0/m_e)}^{1/2}$ and protons $v_{tp}={(k_BT_0/m_p)}^{1/2}$ are $v_{te}=0.06$ and $v_{tp}=1.5 \times 10^{-3}$. The box length is $L_x = 11.84$ along $x$ and $L_y=L_z=L_x/9$ along the other two directions. We resolve $L_x$ with 3600 grid cells and $L_y$ and $ L_z$ with 400 cells each. We use 9 CPs per cell to resolve the protons and electrons, respectively. The ambient plasma is permeated by a magnetic field with the strength $B_0$ that points along the $z$-axis, gives $\omega_{ci}=eB_0/(m_p\omega_{pi}) = 2.1 \cdot 10^{-3}$, and has the pressure $P^* = B_0^2/(2\mu_0n_0k_BT_0)$ ($\mu_0,k_B$: vacuum permeability, Boltzmann constant). 

We inject an unmagnetized electron-positron pair cloud at the left boundary with a mean velocity $v_0=0.6$ along increasing values of $x$. We inject 1 CP per time step for each cloud species at each of the $1.6 \times 10^5$ boundary cells. We give each injected species a non-relativistic velocity distribution with the temperature $50T_0$ and the number density $n_0$ in the simulation box frame. The pair cloud will expand to increasing values of $x$ and its particles will be scattered by their interaction with the ambient plasma. Some will return, cross the boundary where they were injected, and expand into the plasma near the boundary on the other side of the simulation box. Multiple scattering converts directed flow energy into thermal energy and the cloud is close to thermal equilibrium when it crosses the boundary. 

We use $2.6 \times 10^4$ time steps $\Delta_t$ to cover the time $t_{sim}=47$. At this time and due to the periodic boundary conditions, the ambient electrons that were accelerated by the counterstreaming pair clouds start to overlap in the upstream directions of both clouds. The number densities of the overlapping populations are low and they do not drive instabilities. Since $\omega_{ci} \ll \omega_{pi}$ and $\omega_{lh}t_{sim}\approx 0.1$, we can consider the protons to be unmagnetized. The lower-hybrid frequency $\omega_{lh}= {((\omega_{ci}\omega_{ce})^{-1}+\omega_{pi}^{-2})}^{-1/2}$ ($\omega_{ce}=eB_0/m_e$: electron gyro-frequency) is the characteristic frequency of charge density waves in magnetized plasma. Unless stated otherwise, we normalize space to $\lambda_{pi}$, time to $\omega_{pi}^{-1}$, electric fields to $m_p \omega_{pi} c / e$, and magnetic fields to $m_p \omega_{pi} / e$. Table~\ref{table1} gives the box dimensions, simulation time, and value for $B_0$ for several values of $n_0$.
\begin{table}
\begin{tabular}{|c|c|c|c|c|}
\hline
$n_0$ in cm$^{-3}$ & $L_x$ in km & $L_y$ and $L_z$ in km & $t_{sim}$ in msec & $B_0$ in mGauss (10$^{-7}$ T)\\
\hline
$10^{-1}$ & 8500 & 950 & 110 &  0.09\\
\hline
$10^2$ & 270 & 30 & 3.6 & 2.8\\
\hline
$10^5$ & 8.5 & 0.95 & 0.11 & 90\\ 
\hline
\end{tabular}
\caption{Box size, simulation time, and magnetic field amplitude $B_0$ for several number densities $n_0$ of the ambient plasma that surrounds the astrophysical jet.}
\label{table1}
\end{table}

The ion acoustic speed in collisionless plasma corresponds to the hydrodynamical sound speed. Relevant MHD speeds are the Alfv\'en speed along the magnetic field and the fast magnetosonic speed perpendicular to it. In the ambient plasma, the ion acoustic speed is $c_{s} = {((\gamma_e T_e + \gamma_pT_p)k_B /m_p)}^{1/2}$, where $T_e$ and $T_p$ are the electron and proton temperatures. The adiabatic constants in collisionless nonrelativistic plasma are $\gamma_e = 5/3$ for electrons and $\gamma_p = 3$ for protons. The Alfv\'en speed and the fast magnetosonic speed in the ambient plasma are $v_A = B_0 / {(\mu_0m_pn_0)}^{1/2}$ and $v_{fms}={(c_{s}^2+v_A^2)}^{1/2}$. We obtain $c_{s}\approx 3 \times 10^{-3}$ for $T_e = T_p = T_0$, $v_A \approx 2.1 \times 10^{-3}$, and $v_{fms}\approx 3.7 \times 10^{-3}$.

The mildly relativistic speeds of the injected electrons and positrons and their moderate kinetic energies ensure that contributions of radiation reaction processes to the plasma dynamics remain weak compared to collective wave-particle interactions. Pair annihilation and creation are likely to be important on a global jet scale but probably not for the small scales, which we resolve in our 3D simulation. The large grid of the 3D simulation forced us to initialize each plasma species with a low number of particles per cell. This low statistical resolution together with the triangular shape functions of the CPs will cause high amplitudes of statistical electric and magnetic field noise in the ambient plasma. The simulations will, however, not show any significant unphysical particle acceleration caused by this statistical noise. Particle acceleration is caused by the coherent electromagnetic fields driven by the expanding pair cloud. In those box intervals, where the acceleration takes place, the plasma is compressed to several times its initial density. This compression increases the statistical plasma representation in important box intervals. The electrons and positrons of the pair cloud are injected in a y-z plane that is separated by a few cells from the periodic boundary at $x=0$. The drift-Maxwellian is used as the probability distribution function from which the particle velocities are drawn. Particles with the speed $v_0$ cross one cell in about 3 time steps and the injected plasma is thus effectively represented by 3 CPs per cell for electrons and also for positrons. Their interaction with the ambient plasma will compress the pair cloud. The injected pair cloud will eventually be represented by up to 30 CPs per cell for electrons and also for positrons. Since the thermal speed of the injected particles is not much smaller than $v_0$, some of these particles are injected towards the boundary $x=0$. These particles mix rapidly with the pair cloud particles.

\section{Results}

We present the simulation data on a shifted grid. The injection boundary is located at $x=0$ and the pair cloud is injected into the domain $x>0$. The interval $7.4 \le x \le 11.84$ is mapped to $-4.44 \le x \le 0$.  First, we discuss the data that can be compared directly to an MHD model, followed by the detailed analysis of the plasma and field data in the domain $x \le 0$, and conclude with an analysis of that in $x\ge 0$. Volumetric data is rendered with Inviwo~\cite{Jonsson2020}.

\subsection{Connecting structures in MHD and collisionless plasma models}

\begin{figure}
\includegraphics[width=\columnwidth]{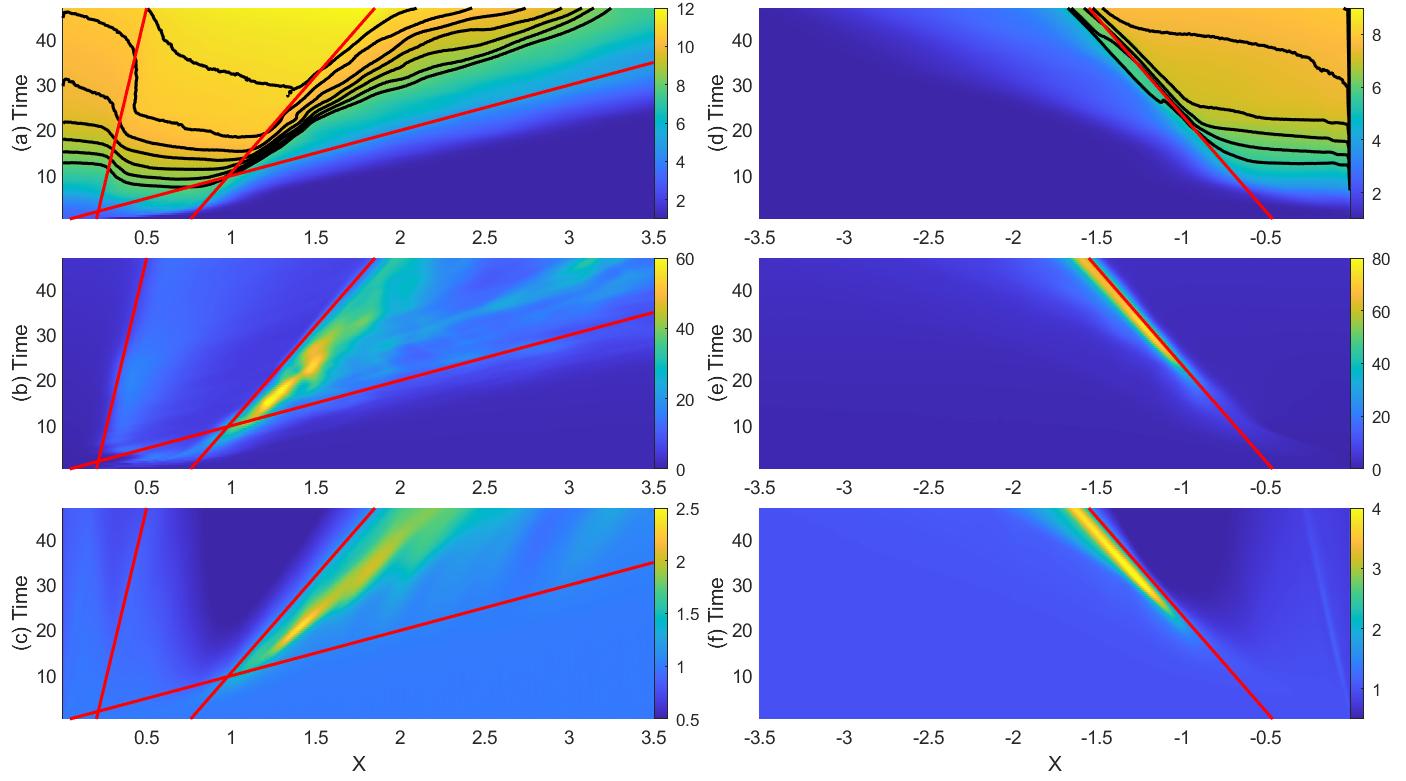}
\caption{Time evolution of the total density of electrons and positrons, of the proton density and of the magnetic pressure $P^*$, which were averaged over $y$ and $z$: The left and right columns show the injected and reflected pair clouds. Panels~(a) and (d) show the sum of the densities of all electron and positron species. The black curves in (a) show the contours 8-11 in steps of 0.5, while those in (d) go from 5.5 to 8.5. In both panels, the contour values increase as we go to larger times. Panels~(b) and (e) show the magnetic pressure. Panels~(c) and (f) show the proton density. All densities are normalized to $n_0$. The red lines in the left column denote the speeds 6.5$\times$10$^{-3}c$, 0.023$c$ (6.2$v_{fms}$), and 0.1$c$. That in the right column corresponds to the speeds -0.023$c$.}
\label{figure2}
\end{figure}

Figure~\ref{figure2} shows the time evolution of the magnetic pressure, the total electron and positron density, and the proton density, which have been averaged over $y$ and $z$. Figures~\ref{figure2}(a)-(c) show three distinctive domains. Near the injection boundary $x=0$, the cloud density is high, the magnetic pressure is low and the proton density is close to its initial value. In Figs.~\ref{figure2}(a) and (b), this domain is bounded by a line from $x\approx 0.2$ at $t=0$ to $x\approx 0.5$ at $t=47$. Its slope corresponds to the speed $\approx 6.5\times 10^{-3}$. To its right, the electron and positron density and magnetic pressure increase, and the proton density decreases. This second domain extends up to the red line that starts at $x\approx 0.75$ at $t=0$ and moves at the speed $0.023c$ or $6.2v_{fms}$ into the ambient plasma. To the right of this line, we find a structure with high magnetic pressure and proton density. Both decrease as we increase $x$ further until the values of the proton density and magnetic pressure become comparable to those of the ambient plasma. The boundary of this third domain at large $x$ moves at a speed $\approx 0.1c$. 

Figures~\ref{figure2}(d)-(f) show the data of the reflected cloud. They reveal two domains. The one to the right of the red line, which denotes the speed $-0.023c$, reveals a fairly uniform electron and positron density with no maximum away from the boundary $x=0$, a low magnetic pressure, and a proton density that decreases with time. To the left of the line, we find a structure with high magnetic pressure and proton density. The proton density peak is located ahead of that of the magnetic pressure at all times. The magnetic pressure, the total electron and positron density, and the proton density decrease as we go farther to the left and converge to the corresponding values of the ambient plasma.

\begin{figure}
\includegraphics[width=\columnwidth]{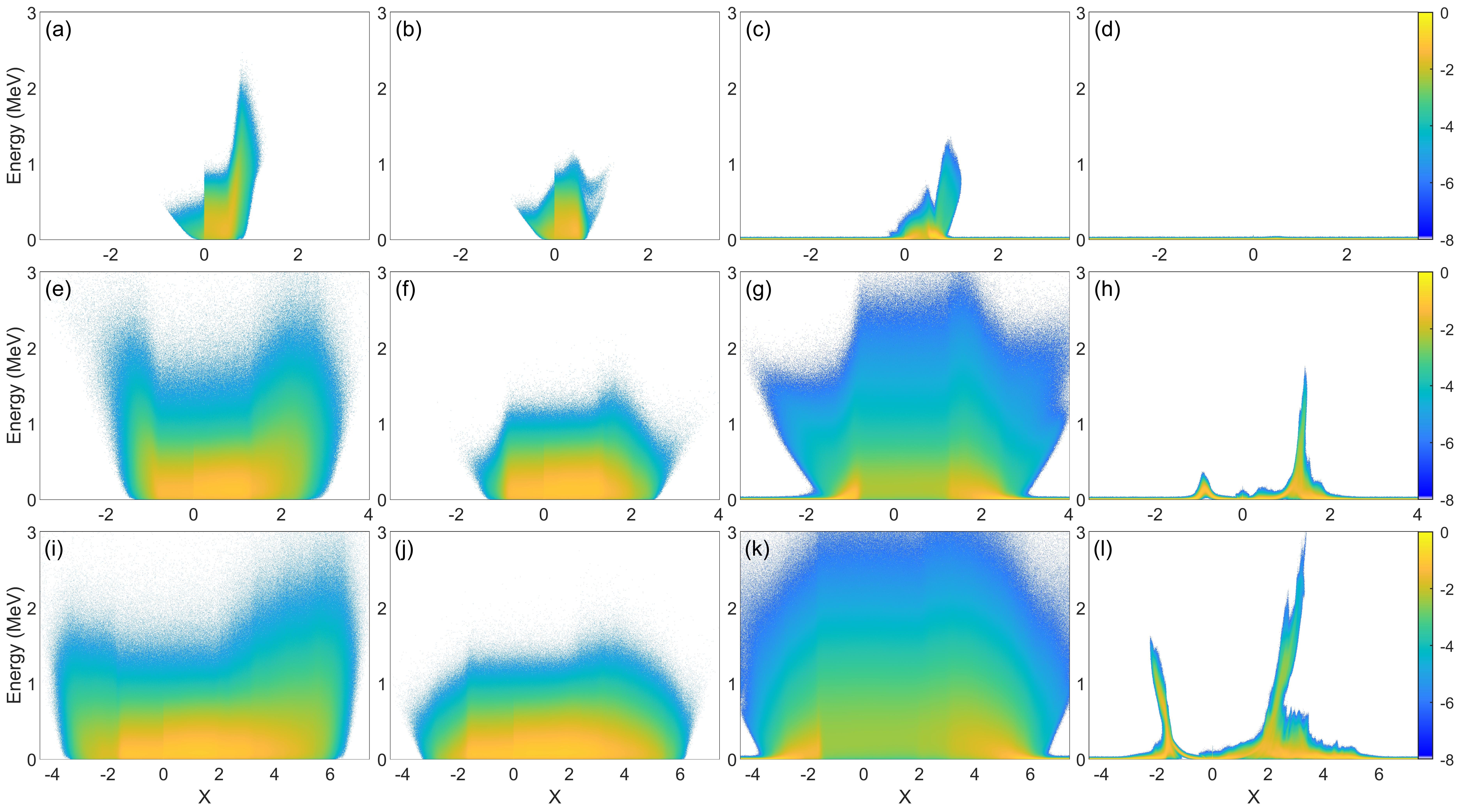}
\caption{Energy distributions of the 4 particle species. Panels (a)-(d) show positrons (left column), cloud electrons (second column), ambient electrons (third column), and protons (right column) at the time $t=1.8$, respectively. Panels (e)-(h) show their distributions at the time $t=18$, while (i)-(l) show those at $t=47$. All densities are normalized to the peak density of ambient electrons at the time $t=0$ and displayed on the same 10-logarithmic color scale.}
\label{figure3}
\end{figure}

Figures~\ref {figure3}(a)-(d) display the energy distributions of the four species at $t=1.8$. Supplementary movie~1 animates them over the time interval $0 \le t \le 47$.
The bulk of the positrons and electrons of the pair cloud are uniformly distributed in the interval $0 \le x \le 0.5$. Some cloud particles have crossed the boundary and entered the domain $x<0$. The larger their energy, the farther they propagate as expected for Larmor rotation in the ambient magnetic field. For comparison, a positron with the energy 340 keV has the relativistic Larmor radius 0.35 in a magnetic field with the strength $B_0$. Positrons are accelerated and cloud electrons are decelerated in the interval $x>0.5$. This effect has been discussed previously~\cite{Dieckmann2018a} and in our section 2.3. If a pair cloud with equally dense positrons and electrons propagates across an electron-proton plasma, instabilities mix the pair cloud particles with the ambient electrons. Since all electrons and positrons have the same modulus of the charge-to-mass ratio, they should reach the same final distribution. However, the electrons are denser than the positrons, and equal speeds of both populations induce an electric field. This electric field accelerates positrons and decelerates electrons in the cloud expansion direction.

At $x=0$, the distributions in Figs.~\ref{figure3}(a,~b) show jumps in the phase space density and in the maximum energy reached by the injected electrons and positrons. At this early time, both jumps are a consequence of the drifting Maxwellian distribution, which we used to initialize the velocities of the injected CPs. The drift velocity $v_0$ is 1.35 times its thermal speed. In the rest frame of this Maxwellian, injected particles cross the boundary if their velocity along $x$ is less than $-v_0$. Injected particles with larger speeds propagate away from the boundary. Around ten times more injected particles move away from the boundary than towards it, which explains the jump in the phase space density. Injected particles, which have a speed $v_0$ along the positive x-axis in the rest frame of the Maxwellian, have a relativistic speed in the box frame. The energy of the ones injected with $-v_0$ is zero; hence injected particles have different maximum energies on both sides of the boundary.  

At the time $t=18$ of the snapshots in Fig.~\ref{figure3}(e)-(h), the spatial interval with uniform electron- and positron distributions has expanded. Positrons reach their highest energies near the fronts of both clouds, while cloud electrons lose energy in this interval. Proton structures have emerged at $x\approx -1$ and $x \approx 1.5$. They are responsible for the density peaks in Fig.~\ref{figure2}(c) and (f). Figure~\ref{figure3}(i)-(l) show the energy distributions at $t=47$. All three electron and positron species have reached a spatially uniform distribution for $-1.5 \le x \le 2$. Positrons carry more energy than the cloud electrons in this interval, while ambient electrons have the largest energy spread. The fastest protons reach the energy 3 MeV, which corresponds to the speed of 0.08. 

\begin{figure}
\includegraphics[width=\columnwidth]{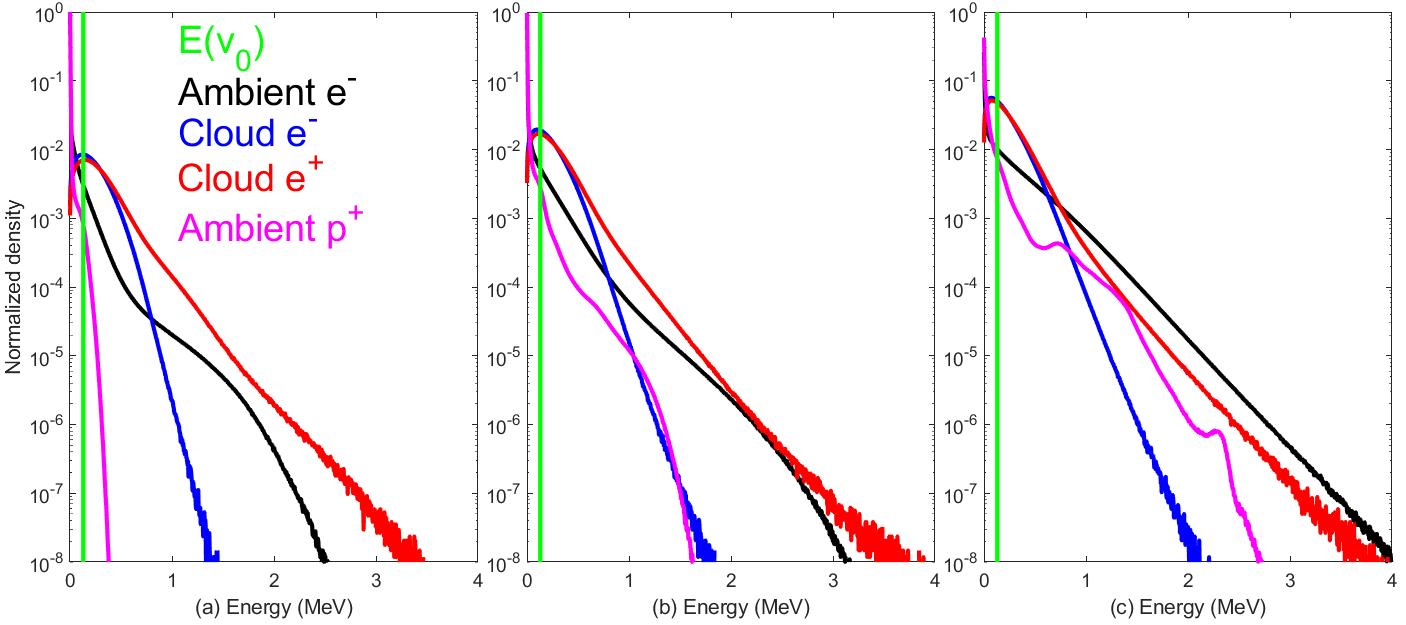}
\caption{Energy distributions of the four particle species, which have been integrated over the full simulation box. Panels~(a),~(b), and~(c) correspond to the times $t=8$, 18, and 47. All curves have been normalized to the maximum value of the ambient electron curve in panel~(a). The green lines denote the energy $E(v_0) =$ 130 keV of an electron moving with the speed $v_0=0.6c$.}
\label{figure4}
\end{figure}

Figures~\ref{figure4}(a)-(c) plot the energy distributions of all species starting at $t=8$.
The energy distributions of the positrons and cloud electrons reach their maxima close to the energy they have if they move with $v_0$. These peak values increase in time due to the permanent injection of pair particles. The energy distributions of cloud electrons and positrons diverge at energies above 1 MeV, which is caused by the positron acceleration and cloud electron deceleration near the cloud fronts in Fig.~\ref{figure3}. Figure~\ref{figure4}(a) shows that ambient electrons were accelerated to energies well above the energy range, which is accessible to cloud electrons. In Fig.~\ref{figure4}(c), the high-energy tails of ambient electrons and positrons decrease at the same exponential rate.

The total electron and positron density, the proton density and the magnetic pressure are variables in collisionless- and MHD plasma and it is interesting to identify structures that we find in both. In particular Figs.~\ref{figure2}(d)-(f) shows a clear separation of the pair cloud from the protons, which were piled up by it. Identifying the mechanisms at work in Fig.~\ref{figure2} will help us understand how collisionless plasmas can sustain MHD  jets. We will resort to a detailed analysis of the PIC simulation data to answer the following key questions: (1)~What is the 3D structure of the pulses of the magnetic pressure and proton density at the front of the reflected pair cloud? (2)~Why are the distributions of the magnetic pressure and proton density diffuse in the domain $x>0$? (3) Why do we have three distinct domains in the domain $x>0$ and two in the domain $x<0$? In what follows, $\langle Q \rangle$ denotes a quantity $Q(x,y,z)$ that was averaged over $y$ and $z$. 

\subsection{Interaction between reflected pair cloud and ambient plasma}

\begin{figure}
\includegraphics[width=0.8\columnwidth]{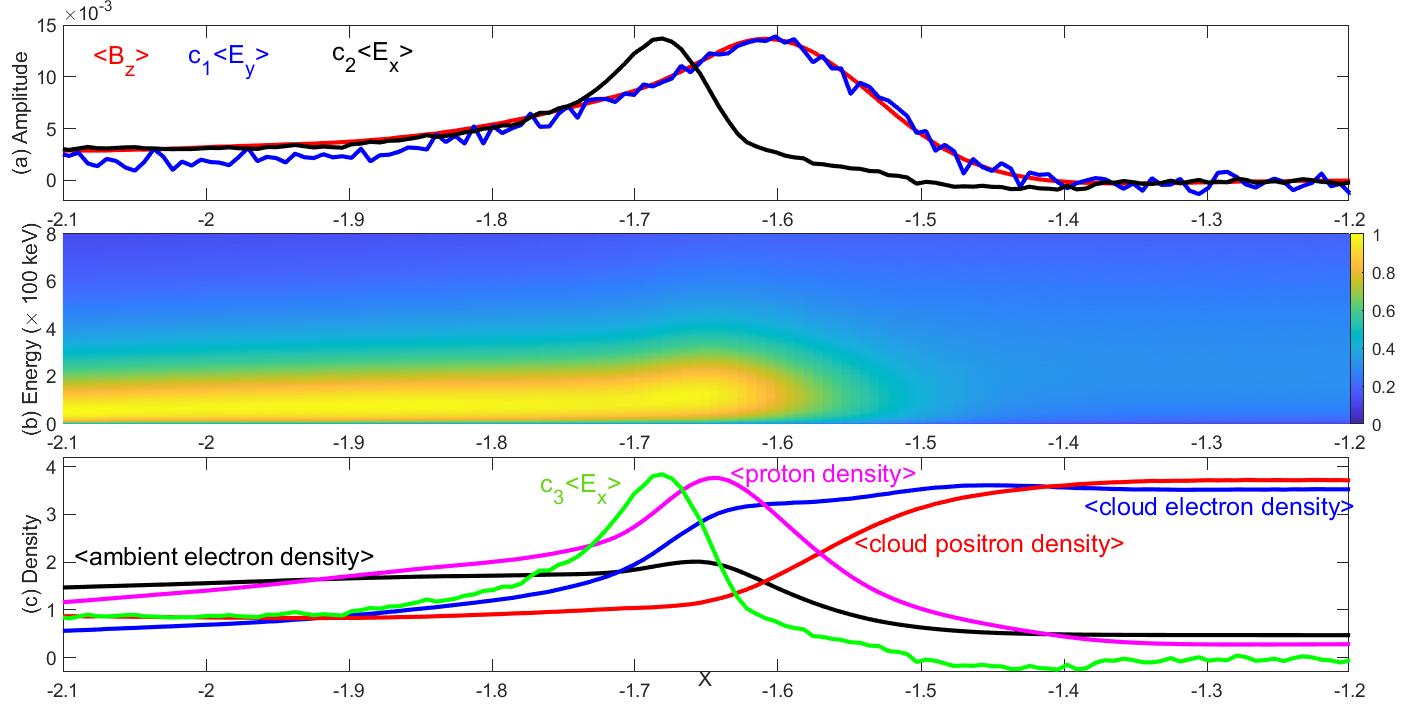}
\caption{Plasma distributions and field amplitudes in the interval $-2.1 \le x \le -1.2$ at $t=47$: Panel (a) shows $\langle B_z \rangle$, $\langle E_y \rangle$, and $\langle E_x \rangle$ with $c_1=-40$ and $c_2=-12.5$. Panel~(b) presents the energy distribution of the ambient electrons near the front of the pair cloud. It is normalized to its peak value. The color scale corresponds to the square root of the phase space density. Panel~(c) plots the density distributions of the 4 plasma species and the amplitude of the electric $E_x$ with $c_3 = -3500$.}
\label{figure5}
\end{figure}

Figure~\ref{figure5}(a) plots $\langle B_z \rangle$, $\langle E_x \rangle$, and $\langle E_y \rangle$ in the interval $-2.1 \le x \le -1.2$.
We do not plot the other field components, because their values are small. The value of $\langle B_z \rangle$ is largest at $x\approx -1.6$, which coincides with the position of the magnetic pulse in Fig.~\ref{figure2}(e). In our normalization, $\langle B_z \rangle$ equals the ratio of the local proton gyro-frequency to the proton plasma frequency. Even at its peak value of $0.014$ or $6.7B_0$, protons would complete only about 10\% of a Larmor orbit until $t=47$; magnetic effects on protons are small. This pulse can, however, confine the bulk of the injected pair plasma since the Larmor radius of an electron with the cloud's mean speed $v_0$ is 0.2 in a magnetic field with strength $B_0$. We find that $\langle B_z \rangle / \langle E_y \rangle \approx 40$ for $x\ge -1.8$. The magnetic pressure pulse moves with the speed $\approx -40^{-1}$ in Fig.~\ref{figure2}(e) and we identify $\langle E_y \rangle$ as the motional electric field of $ \langle B_z \rangle$. Figure~\ref{figure5}(a) also reveals a pulse in the distribution of $\langle E_x \rangle$, which reaches the value $-10^{-3}$ at $x\approx -1.7$ and is still strong at $x=-2.1$. This electric field pulse, which accelerates protons, combined with that in $\langle B_z \rangle$ constitutes an EMP. All fields fluctuate around zero for $x>-1.4$. Absent electromagnetic forces explain why the distributions of positrons and cloud electrons are uniform for $-1.4\le x \le 0$ in Fig.~\ref{figure3}(i) and (j).

In Figure~\ref{figure5}(b), ambient electrons form a compact beam to the left of $x\approx -1.6$ and below 250 keV. Its density, mean energy, and thermal energy peak at $x \approx -1.65$. Electrons with the energy 130 keV, like those of this beam, have a Larmor radius of about 0.04 for $\langle B_z \rangle \approx 0.01$. They are trapped magnetically and pushed forward by the EMP. Trapping means here that the ambient electrons oscillate around a central position $x$ that is phase-locked with respect to the EMP. This works if the magnetic field does not change much across a Larmor radius of the gyrating particle and if the magnetic pulse is large compared to this radius. 

The trapped ambient electrons move with the EMP and their current drives the field $\langle E_x \rangle$, which lets protons and positrons gain energy at the expense of that of cloud electrons. Trapped ambient electrons propagate with the EMP at the speed $-c/40$ and undergo guiding-center drifts. Drift speeds can be estimated for systems that have reached a steady state~\cite{Northrop1961}. Electrons and positrons accelerate faster than protons and at least the trapped electrons and positrons should reach the drift speed. We simplify Eqn.~11 in Ref.~\cite{Rosenbluth1957}, which estimates the drift speed in a nonuniform magnetic field, by assuming that $B_z \gg B_x, B_y$ and that $B_z$ changes only along $x$. The drift speed modulus $|{v}_{y,B}| = (v_{tp}^2/c^2B_z^2) dB_z/dx$ is about $10^{-3}$ for $B_z \approx 0.01$ and $dB_z/dx \approx 0.05$. Another drift mechanism is given by $\mathbf{v}_{EB} = \mathbf{E} \times \mathbf{B}/B^2$ or $v_{y,EB}= -E_x/B_z\approx 0.1$ for $E_x  \approx -10^{-3}$ and $B_z \approx 0.01$. This fastest drift increases the kinetic energy of the trapped ambient electrons by about 2.5~keV, which is well below what is observed in Fig.~\ref{figure5}(b). The trapped electrons have also been heated, which increases the energy where the particle density reaches its maximum value. Given that $v_{y,EB}\approx 2v_{te}$ and that the protons are still at rest, an electron-cyclotron drift instability between ambient electrons and protons will thermalize the beam of drifting ambient electrons~\cite{Forslund1970,Dieckmann2000}. 

According to Fig.~\ref{figure5}(c), the ambient electrons reach their largest density close to where the beam ends in Fig.~\ref{figure5}(b) and it drops to about 0.5 for $x >-1.4$. Figure~\ref{figure5}(b) showed that the EMP could only trap and expel the dense population of ambient electrons with an energy below 250 keV. The energetic ones leaked into the interval $x>-1.6$. Cloud electrons and positrons reach a peak density of about 4 at large $x$ and a detectable number of positrons reach the position $x\approx -2.1$. The proton density increases fastest when $\langle E_x \rangle$ reaches its peak amplitude. Protons reach their maximum density at $x\approx -1.65$, where $\langle E_x \rangle$ decreased to about 25\% of its peak value. The proton density is correlated with $\langle E_x \rangle$. The modulus of $\langle E_x \rangle$ remains high up to $x\approx -2.1$. A comparison with Fig.~\ref{figure3}(i)-(l) ties this electric field to energetic positrons and electrons, which are not dense enough to expel the magnetic field but carry enough current to induce a strong electric field. This electric field grows because the positrons cannot balance the current of the denser electrons. 

\begin{figure}
\includegraphics[width=0.9\columnwidth]{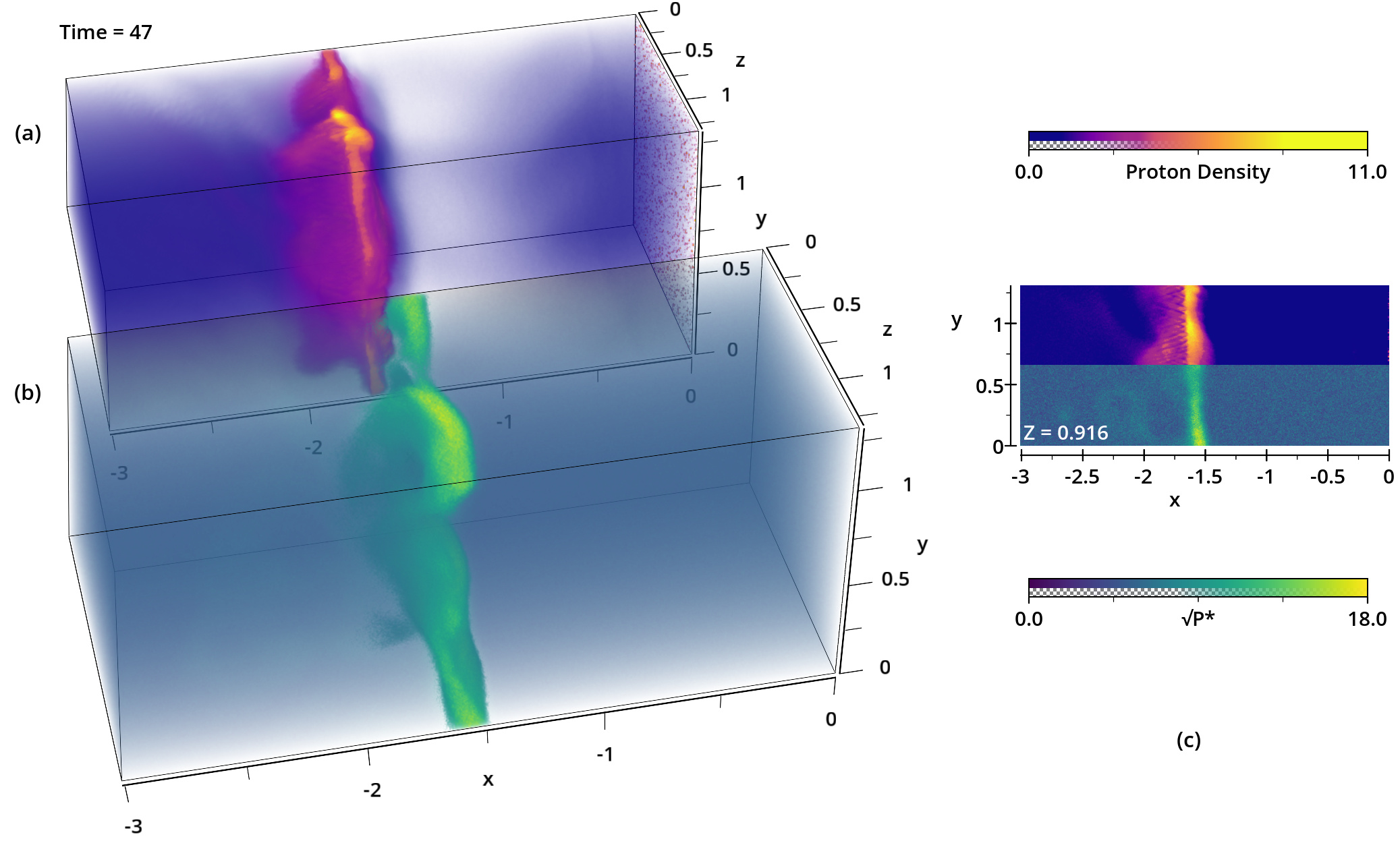}
\caption{Panel (a) renders the proton density in units of $n_0$ and~(b) that of $\sqrt{P^*}$ with the normalized magnetic pressure $P^* = \mathbf{B}^2/(2\mu_0n_0k_BT_0)$. Panel (c) shows data in the slice $z=0.916$. The upper half corresponds to proton density data and the lower one to magnetic pressure. The time is $t=47$.}
\label{figure6}
\end{figure}

Figure~\ref{figure6}(a) and (b) render the pulses in the magnetic pressure and proton density in the domain $x<0$.
Both structures are compact and almost planar and Fig.~\ref{figure6}(c) shows that they are closely connected. The magnetic pressure reaches a peak value that is 250 times larger than its initial one. The pulse in the magnetic pressure trails that in the proton density because the electric field pulse, which reflects the protons, is located ahead of it. There are no density modulations in the interval $-1.5 \le x \le -0.75$; protons have been expelled uniformly from behind the EMP. The supplementary movie~2 animates the data in Fig.~\ref{figure6} in time for $0 \le t \le 47$. The opacity is set such that the initial value 1 for the normalized magnetic pressure is not visible. Magnetic structures move along increasing values of $y$ in time, as expected for a value $v_{y,EB}>0$ in this EMP. According to the supplementary movie~2, their speed along $y$ is initially about 0.2 and decreases as the proton pulse develops. The speed 0.2 exceeds $v_{y,EB}$ but the EMP is neither in a steady state nor perfectly planar. 

\begin{figure}
\includegraphics[width=0.9\columnwidth]{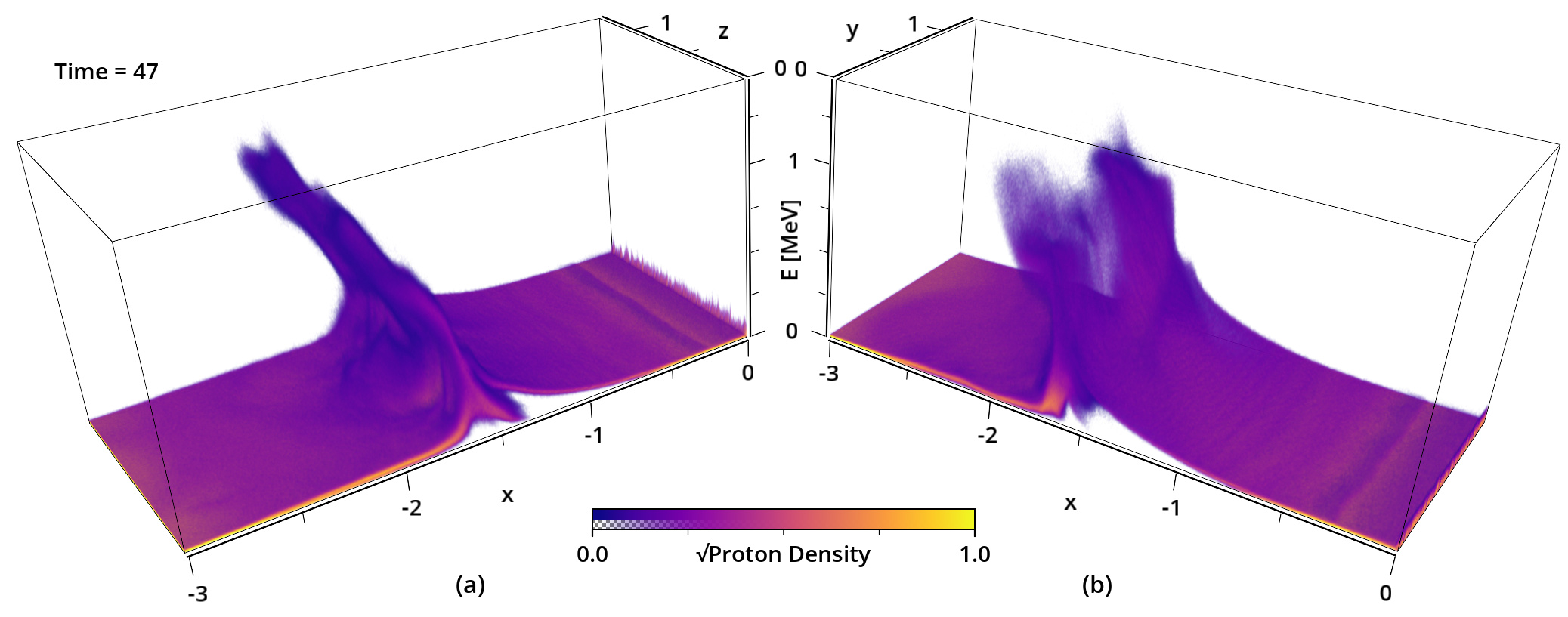}
\caption{Panel (a) and (b) show the energy distributions of protons that interacted with the reflected pair cloud at the time $t=47$. Both distributions are normalized to their peak value at $t=0$ and we take the square root of the phase space density. The distribution in (a) was averaged over $0.63 \le y \le 0.69$ and that in (b) was averaged over $0.63 \le z \le 0.69$.}
\label{figure7}
\end{figure}

Figure~\ref{figure7} shows the proton energy distributions in the x-y and x-z planes. They have been integrated over an interval with the width $0.06$ along the third direction.
Both distributions show that protons are accelerated everywhere along the EMP up to the energy $\approx$ 300 keV of a proton, which moves with the speed of the EMP. The highest energy of about 1.5 MeV is reached by protons, which have been reflected specularly by the EMP. The EMP is not perfectly planar and specular reflection is not always possible, which explains variations of the peak energies with $y$ and $z$. 

Figure~\ref{figure7}(a) and (b) reveal phase space density distributions close to $x=-1.5$ that are double-valued in energy. The upper branch extends to $x=0$ and the energy of its protons decreases with increasing $x$. The supplementary movie~3 animates the energy distributions for $0 \le t \le 47$. It shows that the protons of this branch were accelerated by an electric field, which grew in time while propagating away from the boundary. The second branch of the proton distribution near $x=-1.5$ is connected to the ambient protons at rest and to the protons, which were reflected by the EMP. Protons of this branch have an energy $E\approx$ 300 keV near the beam of reflected protons and they are thus stationary in the rest frame of the EMP. This branch of the proton distribution will eventually become the outer cocoon in the jet model depicted in Fig.~\ref{figure1}. Their mean energy decreases with increasing $x$ and is close to zero at $x\approx -1.4$. Thermal diffusion lets electrons stream from regions with a high proton density to one with a low density like it is the case behind the proton density pulse. An ambipolar electric field grows that points oppositely to the proton density gradient. Downstream protons are accelerated by this field to larger $x$, which slows them down in the simulation box frame. 

\subsection{Interaction between injected pair cloud and ambient plasma}

\begin{figure}
\includegraphics[width=0.9\columnwidth]{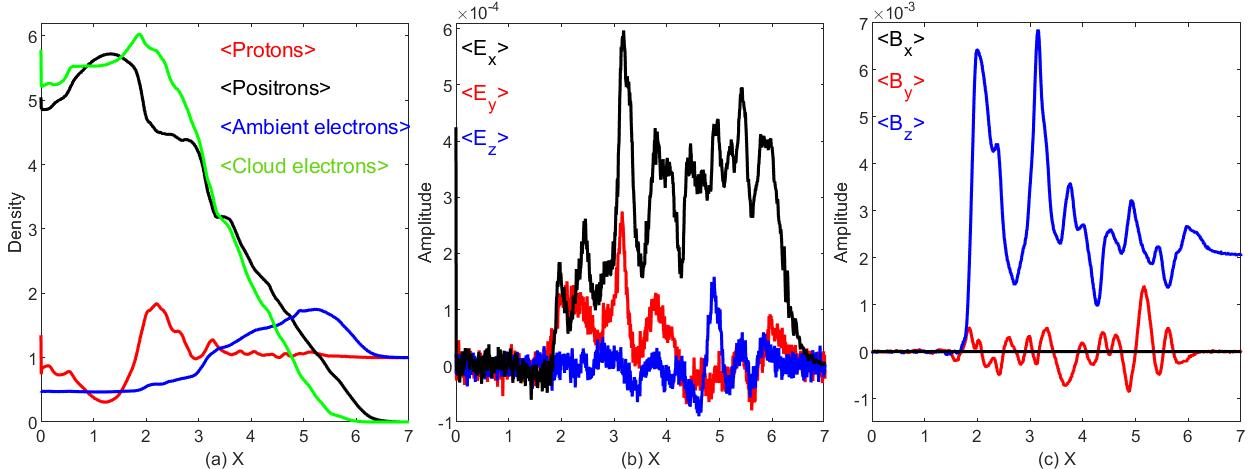}
\caption{Averaged density- and amplitude distributions at $t=47$: Panel~(a) compares the density distributions of all particle species. Panels~(b) and~(c) show the electric- and magnetic field components, respectively.}
\label{figure8}
\end{figure}

In what follows, we address why the distributions in Fig.~\ref{figure2}(a)-(c) are diffuse and why there is a third plasma domain. Figure~\ref{figure8}(a) shows that protons were expelled by the injected pair cloud and piled up ahead of it. Ambient electrons have a density distribution that is qualitatively similar to that in the domain $x<0$. Compared to the reflected pair cloud, the injected one has a higher density at low values of $x$ and its density decreases more slowly with increasing $x$. The field distributions in Figs.~\ref{figure8}(b)~and~(c) are oscillatory and distributed over a wide spatial interval, which demonstrates that they have more internal structure than those in the domain $x<0$. They do not reveal why the magnetic pressure changes at $x\approx 0.5$ in Fig.~\ref{figure2}(b).

\begin{figure}
\includegraphics[width=\columnwidth]{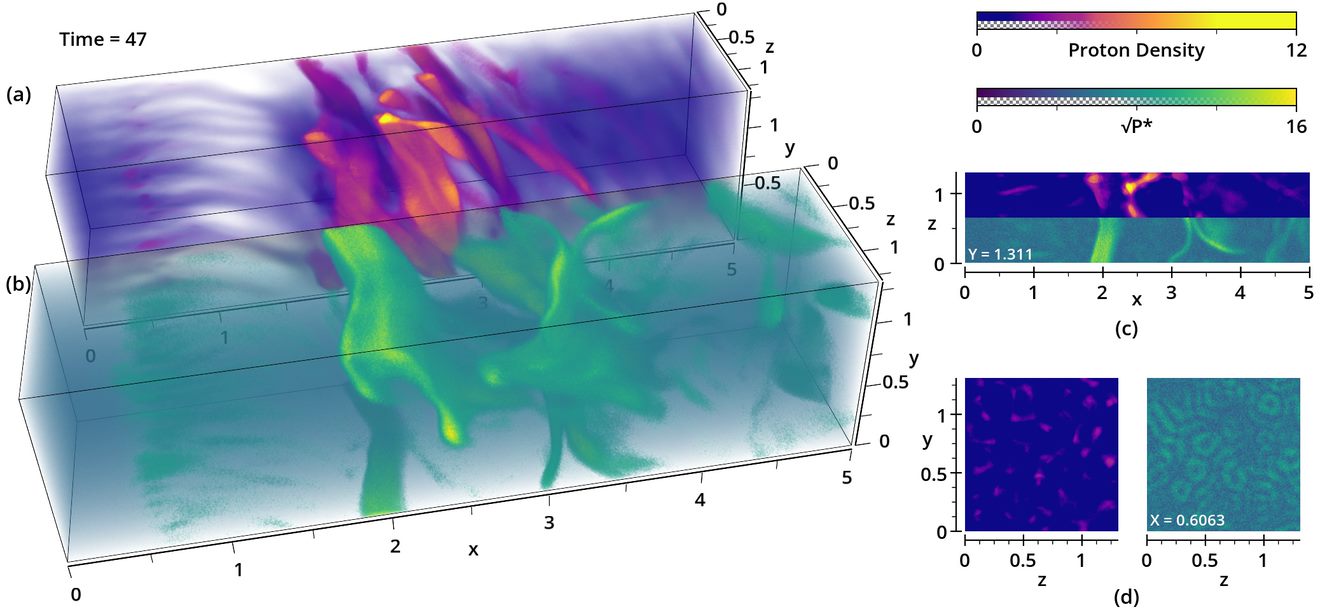}
\caption{Panel (a) renders the proton density in units of $n_0$ and~(b) that of $\sqrt{P^*}$ with the normalized magnetic pressure $P^* = \mathbf{B}^2/(2\mu_0n_0k_BT_0)$. Panel (c) shows data in the slice $y=1.31$. The upper half shows the proton density and the lower half the magnetic pressure. Panel~(d) shows the slice $x=0.6$ of the proton density (left) and magnetic pressure (right). The time is $t=47$.}\label{figure9}
\end{figure}

Figures~\ref{figure9}(a) and (b) render the 3D structure of the pulses in the proton density and magnetic pressure. 
We find several solitary pulses in both distributions for $x\ge 1.8$, which have peak values close to those of the EMP in Fig.~\ref{figure6}. Their electromagnetic fields give rise to the strong oscillations of the box-averaged fields in Figs.~\ref{figure8}(b) and (c). The trailing pulses at $x\approx 2$ in Figs.~\ref{figure9}(a) and (b) suggest that the protons are still accelerated ahead of the magnetic pressure pulse. Both pulses are almost aligned with the y-z plane and move along $x$. At $x=2$, its values $\langle E_y \rangle$ and $\langle B_z \rangle$ in Figs.~\ref{figure8}(b)~and~(c) differ by a factor $40^{-1}$, which equals the propagation speed of the EMP. Like for the EMP ahead of the reflected pair cloud, the electric field $\langle E_y \rangle$ near the trailing EMP at $x\approx 2$ is induced by its moving magnetic field. Correlations between the magnetic and electric fields are less clear at the other EMPs because of their changing orientation and propagation direction. Figure~\ref{figure9}(b) reveals that magnetic field lines are bundled into magnetic flux ropes that follow on average the $z$ direction (See Fig.~\ref{figure8}~(c)) but are twisted around this preferential direction. A missing continuity of the structures in the magnetic pressure and proton density across the split boundary $z=L_z/2$ in Figure~\ref{figure9}(c) underlines that there is no strict correlation between both. Magnetic structures are upheld by electric currents in the dynamic electron and positron flow. A filamentation instability between the injected pair cloud and the ambient plasma~\cite{DieckmannSpencer2020} created current channels, which are enclosed by the tubular structures in the magnetic pressure in the interval $0.5 \le x \le 1.5$. Figure~\ref{figure9}(d) shows their cross-section $x=0.6$. According to Fig.~\ref{figure8}(c), the magnetic field of the tubes averages out to zero in the y-z plane. 

The supplementary movie~4 shows the data, which corresponds to Fig.~\ref{figure9}, for the times $0 \le t \le 47$. The opacity is set such that the initial value 1 for the normalized magnetic pressure is not visible. First, we see magnetic field structures that move rapidly along the negative $ y$ direction. The movie shows that these are magnetic flux ropes that are aligned on average with $z$ while being localized in the $x,y$ plane. We note that magnetic flux rope is a more general term than magnetic flux tube. Reversing the sign of $E_x$ flips the sign of the drift velocity $v_{y,EB}$ compared to that in the supplementary movie~2. The observed slowdown in time of the magnetic structures in the supplementary movies~2 and~4 can be understood as follows. Before the proton reaction, the magnetic field structures are frozen into the drifting electrons and positrons. Protons are not trapped by the EMP and do therefore not remain near the EMP for long enough to be accelerated to a large drift speed. Once the proton density structures have developed in response to the electric field normal to the EMP boundary, they start interacting with the drifting electrons, positrons, and magnetic structures. 

\begin{figure}
\includegraphics[width=0.9\columnwidth]{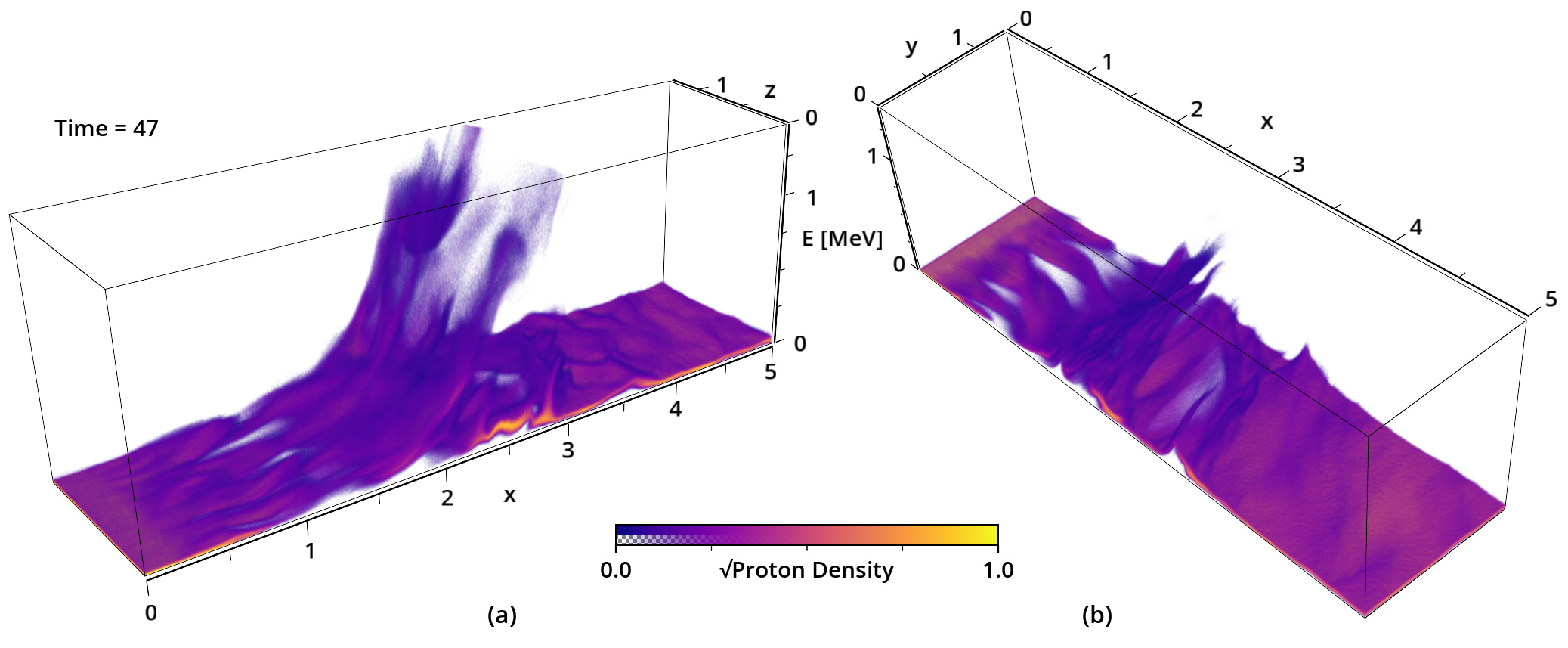}
\caption{Panel (a) and (b) show the energy distributions of protons that interacted with the reflected pair cloud at the time $t=47$. Both distributions are normalized to their peak value at $t=0$ and we take the square root of the phase space density. The distribution in (a) was averaged over $0.63 \le y \le 0.69$ and that in (b) was averaged over $0.63 \le z \le 0.69$.}
\label{figure10}
\end{figure}

Figure~\ref{figure10} visualizes the effects $\langle E_x \rangle$ and $\langle E_y \rangle$ have on the protons by showing their energy distributions along two slice planes. 
Both slices show solitary waves, which are characterized by oscillations of the mean energy. Many protons in the interval $1.8 \le x \le 3.5$ have energies comparable to 300 keV. The EMPs in front of the injected pair cloud are thus able to accelerate protons to their propagation speed $\sim 40^{-1}$. Protons reach higher energies in Fig.~\ref{figure10} than at the front of the reflected pair cloud, which could be caused by interactions with more than one EMP. Since the normals of most EMP boundaries are not parallel to the x-axis, multiple scatterings will accelerate protons also along $y$ and $z$, which increases the number of degrees of freedom accessible to heating. Protons in the interval $x\le 1.8$ are arranged in filaments that are parallel to $x$ and grew because of their interaction with the injected pair cloud. 

The supplementary movie~5 animates the data shown in Fig.~\ref{figure10} in time for $0\le t \le 47$. A perturbation, which accelerates protons, runs at the speed $\approx$ 0.1 (See Fig.~\ref{figure2}(a)-(c)) to increasing $x$ while drifting along $y$. According to Fig.~\ref{figure4}(c), no proton reaches the energy $\approx$ 5 MeV that would correspond to a drift speed $v_{y,EB}=0.1$. Hence, the drift motion of the perturbation in the proton distribution is due to their interaction with the electromagnetic fields of the EMPs that convect with the drifting electrons and positrons.  

\begin{figure}
\includegraphics[width=0.9\columnwidth]{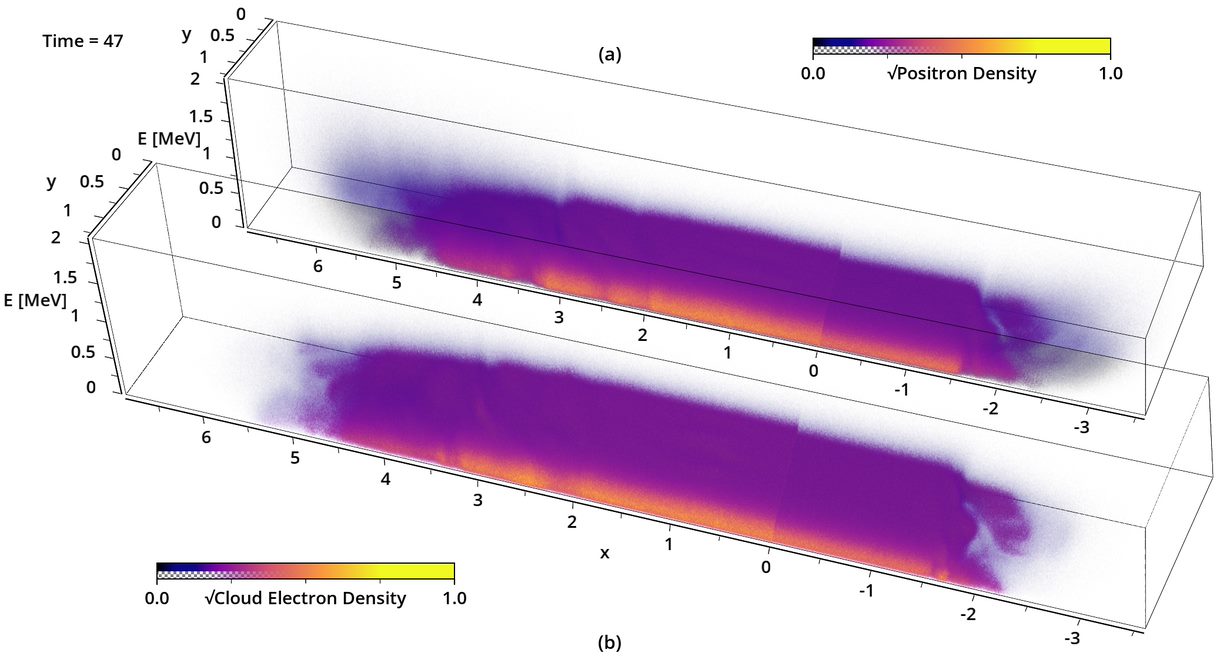}
\caption{The energy distributions of the cloud electrons and positrons at $t=47$, which were averaged over $0.63 \le z \le 0.69$. The upper volume (a) renders that of the positrons and the lower one (b) that of the cloud electrons. Both phase space densities are normalized to the peak value of that of ambient electrons at $t=0$. The color scale denotes the square root of the densities.}
\label{figure11}
\end{figure}

Figure~\ref{figure11} renders the energy distributions of positrons and cloud electrons and the supplementary movie~6 tracks their evolution during $0 \le t \le 47$. 
Electrons and positrons have a spatially uniform energy distribution over the interval $-1.5 \le x \le 2$. Those in the interval $x>0$ have a larger overall density, which explains the apparent jump near $x=0$ of the otherwise uniform distribution. The distributions of the cloud electrons and positrons stop being uniform at the position $x \approx -1.5$ of the EMP ahead of the reflected pair cloud and in the interval $x\ge 2$ that is filled with several EMPs.  

\begin{figure}
\includegraphics[width=0.9\columnwidth]{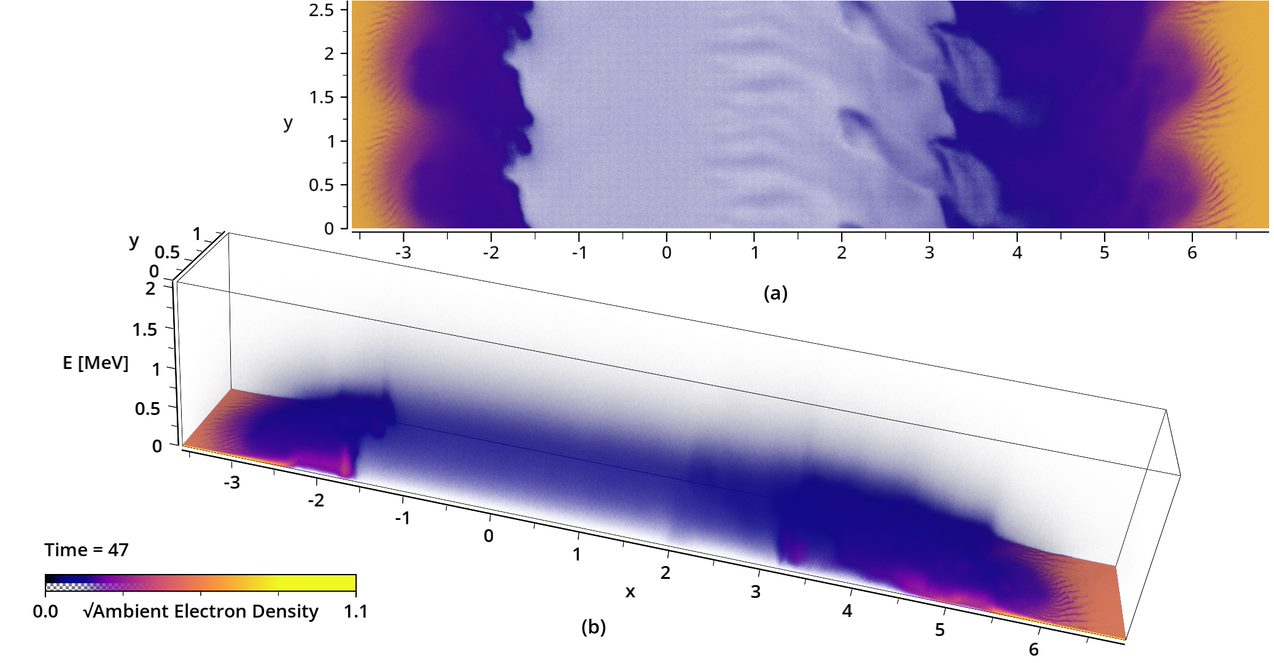}
\caption{The energy distribution of the ambient electrons at $t=47$, which was averaged over $0.63 \le z \le 0.69$ and shown from two view directions. We duplicated the distribution in panel~(a) along $y$ to make it easier to track structures across the periodic boundary at $y\approx 1.3$. The distribution is normalized to its peak value at $t=0$ and the color scale denotes the square root of the density.}
\label{figure12}
\end{figure}

Figure~\ref{figure12} shows the energy distribution of the ambient electrons. 
It is hot and diffuse for $-1.5 \le x \le 2$. Density structures in the interval $0.5 \le x \le 1.5$ can be tied to those in the proton density and magnetic pressure, which we associated previously with a filamentation instability between protons and the injected pair cloud. Ambient electrons have a much higher density ahead of the EMP at $x=-1.5$ and in the region $2 \le x \le 6$. This larger density implies that the distribution in Fig.~\ref{figure12} is denser too. Stripes can be seen at $x\approx 6$ and $x\approx -3.2$. On average, their wavevector is aligned with $y$. The perturbations follow the front of the ambient electrons and their wavelength along $y$ is about $\approx 0.1$ in front of the injected and reflected pair clouds. 

They are caused by an electron-cyclotron drift instability between drifting ambient electrons and protons at rest. It leads to the growth of density waves. Their exponential growth rate is comparable to $\omega_{pi}$ and they yield oscillations according to the resonance condition $\omega_u/k_u = v_D$, where $\omega_u$ and $k_u$ are the frequency and wavenumber of the growing wave in the rest frame of the drifting electrons and $v_D$ is the drift speed between electrons and protons. If we assume that $\omega_{u}=\omega_{pe}$ ($\omega_{pe}=\sqrt{1836}\, \omega_{pi}$: electron plasma frequency) the resonance condition gives a wavelength $\lambda_u = 2\pi v_{y,EB}/\sqrt{1836} \approx 0.015$. The wavelength of such waves can become larger when they saturate nonlinearly and we can also have a larger drift speed than $v_{y,EB}$, which was based on $\langle E_x \rangle$ and $\langle B_z \rangle$.

The supplementary movie~7 tracks the evolution of the energy distribution shown in Fig.~\ref{figure12} for $0 \le t \le 47$. It confirms that the structures in the distribution of trapped electrons move in opposite directions for $x<-1.5$ and $x>2$. The stripes in the density at the front of the EMP in the interval $x<0$ and at the front of the leading EMP in the interval $x>0$ are stable and hardly move along the EMP boundary; electron-cyclotron waves propagate only slowly in the rest frame of the protons.

\section{Discussion}

Our simulation setup allows us to study self-consistently the growth and evolution of discontinuities between an electron-positron pair plasma and ambient plasma, which consists of electrons and protons. We focused here on interactions between a cloud of unmagnetized pair plasma and an ambient plasma, which was permeated by a magnetic field that was oriented orthogonally to the mean velocity vector of the injected cloud. The injected pair cloud had a mean speed of 0.6$c$. It expelled the ambient magnetic field of the ambient plasma and piled it up at its front. Ambient electrons were trapped by this magnetic pulse and pushed across the protons. Their electric current led to the growth of an electric field pulse just ahead of the magnetic one that accelerated the protons into the expansion direction of the pair cloud.

Behind this EMP, the injected pair cloud interacted with the now unmagnetized protons via a filamentation instability. Its magnetic field rearranged pair cloud particles into current flux tubes that were aligned with their initial mean velocity vector. Protons were accelerated by electric fields and compressed into density filaments. Due to this plasma rearrangement, the pressure imposed by the pair cloud on the EMP varied as a function of the position on the EMP boundary. Pair cloud particles were scattered by the electromagnetic fields behind the EMP and some crossed the periodic boundary of our simulation; a reflected pair cloud expanded into the ambient plasma at the other end of the simulation box. Its mean speed was too small to drive a filamentation instability and the pressure the reflected cloud imposed on the EMP was uniform.

The EMPs could not expel all protons. Concerning the jet in Fig.~\ref{figure1}, this means that a diluted proton population remains in its spine. A filamentation instability between them and the outflow will slow down and heat the latter. Increasing the mean speed of the injected pair cloud in our simulation or that of the outflow in Fig.~\ref{figure1} beyond 0.6$c$ is unlikely to change this because filamentation instabilities get stronger with increasing mean speeds. This was demonstrated by a previous PIC simulation~\cite{Spitkovsky2008}, where a filamentation instability between two electron-ion plasmas could mediate a highly relativistic shock. The directed flow speed of the pair outflow in box F in Fig.~\ref{figure1} will thus be transferred to protons behind the EMP and heat the outflow before it reaches the discontinuity. The thermal pair plasma, which drives the lateral expansion of the jet (Box R), will expel protons without driving filamentation instabilities as in the case of our reflected pair cloud. In both cases, the pair plasma next to the discontinuity is hot and slow as expected for the plasma of the inner cocoon.

In its rest frame, the forces imposed on the EMP by the pair plasma and the ambient plasma cancel each other out. The thermal pressure the injected pair cloud had after the filamentation instability was comparable to that of the reflected one. Hence, the EMPs ahead of both pair clouds expanded at approximately the same speed, which was about 6.2 times the fast magnetosonic speed in the ambient plasma. Initially, the thermal pressure of the pair cloud is balanced by the ram pressure of protons. We saw the growth of an outer cocoon behind the EMP that was located ahead of the reflected pair cloud in Fig.~\ref{figure7}(a) near $x\approx -1.5$. The proton structure to the right of the reflected protons will eventually become the downstream region of the shock, which moves to the left and reflects some of the ambient protons. Once this outer cocoon has formed, its thermal pressure will replace the proton ram pressure as the means to balance the pressure of the inner cocoon. 

Changes in the magnetic field amplitude across the EMP require oppositely directed electric currents ahead and behind the EMP. The latter is the diamagnetic current at the surface of the unmagnetized pair cloud. Its high temperature lets the drift speed of its electrons and positrons be small compared to their thermal speed, which reduces the growth rate and impact of drift instabilities. Ahead of the EMP, the electric current is caused by trapped ambient electrons drifting in the EMP. We observed in Fig.~\ref{figure12} density waves in the energy distribution of ambient electrons ahead of the leading EMPs at $x\approx 6.2$ and $x\approx -3.2$. Their short wavelength and rapid growth rates tie them to an electron-cyclotron drift instability between the drifting ambient electrons and protons at rest. The wavevector of the unstable waves is parallel to the drift velocity vector of the ambient electrons. This direction was not resolved in the 2D simulation in Ref.~\cite{Dieckmann2020b} and the ambient electrons remained cool. Resolving this direction heated the ambient electrons to a relativistic temperature and gave them an exponentially decreasing energy spectrum like the one we observe here~\cite{Dieckmann2022}. 

Drift waves have a wavevector that is parallel to that of the interchange mode of the RT-type instability of the EMP. Drift instabilities are thought to accelerate the growth of the interchange/ballooning instability of the closely related tangential discontinuity~\cite{Pritchett2010}. In the supplementary movie 7, we observed throughout the simulation similar structures in the density of the trapped ambient electrons at the front of the injected and reflected pair cloud. The EMP ahead of the reflected pair cloud showed only weak surface deformations, which indicated the onset of an RT-type instability at late times. In contrast, the initial EMP ahead of the injected pair cloud collapsed almost instantly and gave way to a transition layer filled with several EMPs. This transition layer developed as fast as the one ahead of the injected pair cloud in the 2D simulation that resolved the interchange mode~\cite{Dieckmann2022}. We may attribute this collapse, which progressed faster than expected from theory~\cite{Winske1996} to the spatially non-uniform pressure of the injected pair cloud. The lifetime of the EMP ahead of the reflected pair cloud exceeded by far that in the 2D PIC simulation of the interchange mode. We attribute this to a competition between the interchange- and undular modes. Growing undular modes break the alignment of the wave vector of the interchange mode with the flow direction of the drift current and, hence, the wavevector of the seed perturbations. Their decoupling may reduce the growth rate of the coupled instability from that of the drift instability to that of the RT interchange mode. 

An initially planar magnetic boundary like the EMP is transformed by the interchange instability into magnetic flux ropes, which are twisted by the undular mode. Indeed, the renderings of the magnetic pressure ahead of the injected pair cloud showed that the magnetic field arranges itself into 3D structures that can loosely be described as magnetic flux ropes. These structures grow and evolve in the transition layer between the injected pair cloud and the ambient plasma and their enormous magnetic pressure can transfer energy from the pair cloud to protons. The magnetic flux ropes will also interact with the uniform magnetic field of the ambient plasma and their complex and changing three-dimensional shape can initiate reconnection in an MHD model~\cite{Cargill1996}. Magnetic reconnection in collisionless plasma~\cite{Guo2014,Melzani2014} can create distributions of electrons and positrons that follow a power law as a function of particle energy. In our simulation, we observed an exponential decrease in the distribution of electrons and positrons with energy, which might be caused by the relatively short simulation time or by a low relativistic factor of particles. 

The front of the transition layer ahead of the pair cloud, which we injected at the speed 0.6$c$, expanded at the much lower speed 0.1$c$ into the ambient plasma. This layer is thus capable of slowing down and compressing a pair plasma outflow, which is what we require from a discontinuity. Its width grew to a few proton skin depths at the end of the simulation, which was also observed in the 2D simulations~\cite{Dieckmann2020b,Dieckmann2022}. 

The densities of pair plasma jets and their ambient medium depend on physical conditions near the accreting compact object, which can vary drastically in time and for different accreting compact objects. Let us assume the transition layer propagates through the wind of our sun somewhere between its surface and  Earth. The solar corona has a number density $n_0 \approx$ 10$^9$ cm$^{-3}$ and the solar wind near the Earth has a value $n_0 \approx$ 10 cm$^{-3}$. For $n_0 \approx$ 10$^5$ cm$^{-3}$, a thickness of 3 proton skin depths corresponds to about 2 km and the peak amplitude of the EMP $\approx 15B_0$ is about $10^{-4}$ T. In the corona, the thickness decreases by a factor of 100, and the magnetic field increases by the same factor. The opposite is true for $n_0 \approx$ 10 cm$^{-3}$. Even a thickness of 200 km would be far less than the diameter of astrophysical jets and many orders of magnitude less than the mean free path of charged particles in the solar wind.

We resolved the ambient plasma and the injected pair plasma by a relatively low number of particles per cell. Given that the important wave-particle interactions took place in regions with a plasma density that exceeded the initial one by a factor of 3 and more, the statistical plasma representation was high enough to resolve important structures like the flux ropes and the drift waves ahead of the EMPs. We chose a grid cell size that was about twice the electron Debye length of the ambient plasma in order to maximize the volume of the simulation box and the simulation time step. For the triangular shape function we used for the computational particles, this cell size can result in moderate self-heating of the plasma~\cite{Arber2015}. During our short simulation time, we could not observe significant self-heating of electrons and protons. Both species maintained a thermal speed that was low compared to that of the injected pair plasma; we could consider the ambient plasma to be cold. Future PIC simulations with a larger number of particles per cell and a finer grid may reveal processes we could not observe here, but such simulations are currently too expensive to perform.

\section*{Acknowledgements}

The simulations were performed on resources provided by the Swedish National Infrastructure for Computing (SNIC) at the NSC and on the centers of the Grand Equipement National de Calcul Intensif (GENCI) under grant number A0090406960. The first author also acknowledges financial support from a visiting fellowship of the Centre de Recherche Astrophysique de Lyon.

\end{document}